# Personalized Driving Behaviors and Fuel Economy over Realistic Commute Traffic: Modeling, Correlation, and Prediction

Yao Ma, *Member, IEEE* and Junmin Wang, *Senior Member, IEEE*

*Abstract*— Drivers have distinctively diverse behaviors when operating vehicles in natural traffic flow, such as preferred pedal position, car-following distance, preview time headway, etc. These highly personalized behavioral variations are known to impact vehicle fuel economy qualitatively. Nevertheless, the quantitative relationship between driving behaviors and vehicle fuel consumption remains obscure. Addressing this critical missing link will contribute to the improvement of transportation sustainability, as well as understanding drivers' behavioral diversity. This study proposed an integrated microscopic driver behavior and fuel consumption model to assess and predict vehicle fuel economy with naturalistic highway and local commuting traffic data. Through extensive Monte Carlo simulations, significant correlation results are revealed between specific individual driving preferences and fuel economy over drivers' frequent commuting routes. Correlation results indicate that the differences in fuel consumption incurred by various driving behaviors, even in the same traffic conditions, can be as much as 29% for a light-duty truck and 15% for a passenger car. A Gaussian Process Regression model is further trained, validated, and tested under different traffic and vehicle conditions to predict fuel consumption based on drivers' personalized behaviors. Such a quantitative and personalized model can be used to identify and recommend fuel-friendly driving behaviors and routes, demonstrating a strong incentive for relevant stakeholders.

*Index Terms*—Driver Behavior, Fuel Economy.

## I. INTRODUCTION

WITH the rapid development of the automotive industry and increasing vehicle ownership, the number of motorized vehicles have topped one billion worldwide as of 2010. While the growing availability of vehicles has greatly facilitated personal and commercial transportation needs, it also brings about critical challenges, including energy consumption and pollution [1]-[3].

Significant efforts have been dedicated to improving the efficiency and sustainability of the ground transportation system from different perspectives, including vehicle technology advancements, enhanced infrastructures, and proper traffic policies. One factor that is often overlooked, though, is the human factor and its impacts on the vehicle fuel economy in real traffic scenarios. This study will quantitatively investigate this issue and demonstrate tangible benefits through common and realistic traffic case studies.

Existing literature has investigated the significance of driving behaviors primarily for safety enhancement purposes [4], considering a human driver inevitably encounters distractions [5], reaction time [6], and operating mistakes [7] that lead to the deterioration of driving performance. Therefore, it is crucial to understand how human driving preferences affect vehicle operation through driver behavior modeling [8]. In [9], the authors describe the characteristics of human drivers and provide in-depth guidance on modeling approaches. Physical limitations and attributes like human delay, visual and motion influences, and preview utilization are strongly suggested when modeling human driver behaviors. In [10], authors use a Markov dynamic model to infer drivers' intended actions based upon observed temporal patterns of environmental and behavior state. Simulator studies under various driving conditions, including emergency maneuvers, show the effectiveness of this approach.

Drivers often have unique driving preferences attributed to diverse demography, mental state, travel purposes, among other factors. Such diversity directly translates to different vehicle maneuvers even facing the same traffic condition [11], resulting in different vehicle motion dynamics that substantially impact fuel consumption. In [12], the authors investigate the impacts of the human driver on an intelligent transportation system from the perspective of traffic efficiency and average travel speed. Probabilistic models are shown capable of modeling driver behavior based on empirical measurements. In [13], the authors report a long-term training program for urban bus drivers towards more fuel-efficient driving behaviors. It is found that while the fuel-saving effects are strong during the training period, such improvement does not translate to drivers working situations. In [14], the authors investigate the effectiveness of providing drivers with road information to reduce traffic congestion. Under the assumption that drivers have perfect information about road capacity, travel costs are reduced as expected. However, when imperfect information is delivered to drivers, the reduction is absent.

For a given vehicle, in addition to various driving behaviors, the fuel consumption is also heavily influenced by traffic conditions such as traffic density and travel time [15]. Therefore, the exact characterization of the relationship between driving behaviors and fuel economy in real-world

This work was supported by the National Science Foundation under Grant No. 2153229. Yao Ma is with Department of Mechanical Engineering at Texas Tech University, Lubbock, Texas, 79409, USA (Email: yao.ma@ttu.edu);

Junmin Wang is with Walker Department of Mechanical Engineering at The University of Texas at Austin, Austin, Texas, 78705, USA (Email: JWang@austin.utexas.edu).

driving scenarios becomes highly challenging due to traffic uncertainties. In this study, we mitigate such uncertainties by limiting our scope to daily commuting traffic, where we continuously collect the daily commuting traffic data over a certain number of frequent routes [16]. Since the daily traffic patterns are similar on the frequent routes during the same travel time, the impact of traffic uncertainty is reduced. This is based on a reasonable assumption that drivers often operate vehicles over frequent daily routes, considering many drivers' commute routes are limited [17]. In addition, most commercial fleets, such as parcel delivery, urban public transportation, and shared mobility service operate according to specific routes and schedules on a daily basis. Hence the revelation of the relationship between driving behaviors and vehicle fuel consumption can have potentially significant economic incentives for the logistics and transportation sectors.

An accurate and computationally friendly driving behaviors model and fuel consumption estimation are indispensable to examine the quantitative correlation between driving behaviors and vehicle fuel economy. In [18], a continuous microscopic traffic model (Intelligent Driver Model) is developed and calibrated with experimental traffic trajectories. The model describes the longitudinal vehicle dynamics as a function of both traffic conditions (desired freeway speed, speed limit, relative speed, and distance) and driving behaviors (acceleration and braking aggressiveness, preview time horizon) in a deterministic formulation. The model shows realistic driving behaviors and produces no collision. Besides the driving model, a precise estimation of vehicle fuel consumption is critical for high fidelity results. A desirable fuel consumption model should be both accurate and generalizable. In [19], a power-based vehicle fuel consumption model is proposed to predict the instantaneous fuel rate using vehicle acceleration and speed. The model is validated against experimental data and demonstrates a high level of accuracy. The model can be calibrated with publicly available data.

This study aims to quantitatively characterize the impacts of individual driving behaviors on vehicle fuel economy over frequent routes in real-world traffic scenarios. Such a characterization enables the potential of predicting and assessing vehicle fuel economy by observing human driving preferences through daily operations such as commuting, delivery, and public transportation. An accurate characterization of the relationship between driving behaviors and fuel economy can be particularly advantageous in managing large commercial fleets by guiding drivers towards cost-effective behaviors and routes. It can also facilitate the development of Advanced Driver-Assistance Systems (ADAS) to better cope with personal preferences that will improve drivers' comfort and trust. By exploiting the growing availability of human driving data, such benefits can be anticipated in the foreseeable future.

The contributions of this study include:
1. A computational-efficient microscopic driving model with fuel consumption estimation is constructed based on naturalistic traffic trajectories.
2. Statistically significant correlations between drivers' diverse behaviors and vehicle fuel economy are revealed through extensive Monte Carlo simulations.
3. A prediction model is trained, validated, and tested to predict vehicle fuel consumption based on human driving preferences with high accuracy and generalizability.

The rest of this paper is organized as follows. In section II, the integrated driving behavior and fuel consumption model is established and calibrated. In section III, the numerical correlations between driving preferences and fuel economy are investigated with experimentally collected traffic data during local and highway commuting. Simulation results are presented and discussed. Concluding remarks are made in the last section of the paper.

## II. DRIVING BEHAVIOR AND FUEL CONSUMPTION MODEL

A dynamic driving behavior model integrated with vehicle fuel consumption estimation is developed in this section. The proposed model consists of two submodules, namely a microscopic driving model to describe the individual driving behaviors in realistic traffic and a vehicle fuel consumption model to provide an instantaneous estimation of fuel consumption rate. The structure of the proposed model is shown in Figure 1.

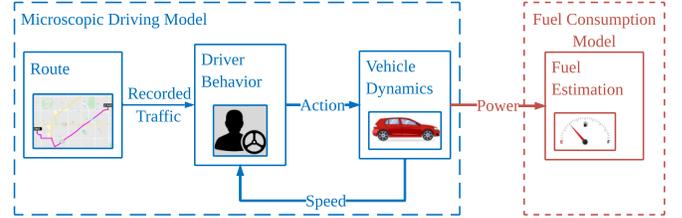

Figure 1. Driving behavior and fuel consumption model

### A. Microscopic Driving Model

The driving behavior is modeled by the Intelligent Driver Model (IDM) as introduced in [18]. The IDM model characterizes longitudinal vehicle dynamics as in (1) and (2),

$$\dot{v}_i = a_i \left( 1 - \left( \frac{v_i}{v_{0_i}} \right)^\delta - \left( \frac{s^*(v_i, \Delta v_i)}{s_i} \right)^2 \right), \quad (1)$$

$$s^*(v_i, \Delta v_i) = s_{0_i} + T_i v_i + \frac{v_i \Delta v_i}{2\sqrt{a_i b_i}}, \quad (2)$$

where $v$ is vehicle velocity; subscript $i$ represents the $i$th vehicle in the traffic; $a$ is the maximum acceleration; $v_0$ is the desired freeway speed; $\delta=4$ is the acceleration component; $s$ is the actual gap distance; $s^*$ is the desired minimum gap distance; $s_0$ is the jam distance; $T$ is the time headway; $\Delta v$ is the speed difference to the leading vehicle; $b$ is the desired deceleration.

As shown in (1), the longitudinal vehicle dynamics is a nonlinear function of multiple coupling factors that encodes both traffic conditions and driving behaviors. For the sake of conciseness, we briefly discuss this under two typical traffic conditions. First, when a vehicle is traveling at high speed in light traffic, the actual gap distance $s$ is much larger than the desired minimum gap distance $s^*$. The vehicle acceleration is mainly determined by the desired freeway speed and actual vehicle speed since $s^*/s$ approximates to zero. As vehicle speed approaches the desired freeway speed, the right-hand side of Eq. (1) converges to zero. Therefore, the desired freeway speed is reached and maintained. Second, when a vehicle is traveling at a low speed in heavy traffic, the acceleration is mainly determined by the desired gap distance and actual distance since the vehicle speed is much lower than the desired freeway speed and $v/v_0$ approximates to zero. In this scenario, the car-following policy tries to regulate the actual gap distance to the desired gap distance. When the desired distance is reached, the acceleration reduces to zero, and speed is maintained. These two trivial cases demonstrate the IDM working mechanism in general. A rigorous analysis of stability can be found in [18]. The traffic conditions in practice are often a mixture of the two scenarios mentioned above.

In addition to traffic conditions, human driving preferences also affect longitudinal vehicle dynamics significantly. In the IDM, the driving preferences are characterized by several intuitive parameters with physical implications. For instance, an aggressive driver may apply throttle harder and result in large maximum acceleration. On the other hand, a cautious driver may preview a longer time headway to avoid unnecessary speed fluctuations. Under heavy traffic conditions, the desired minimum gap distance also varies among different drivers. These examples show that driving behaviors can explicitly affect vehicle dynamics on the microscopic level, and such effects are well parameterized in the IDM model.

*B. Vehicle Dynamics*

The instantaneous vehicle power is calculated as in (3) [22],

$$P(t) = \frac{v(R(t) + 1.04m\dot{v})}{3600\eta_d}, \quad (3)$$

where $v$ (m/s) is vehicle speed; $m$ (kg) is vehicle mass; $\eta_d$ is driveline efficiency; $R(t)$ (N) is resistance force as in (4),

$$R(t) = \frac{\rho}{25.92} C_D C_h A_f v^2 + mg \frac{C_r}{1000}(c_1 v + c_2) + mgG, \quad (4)$$

where $\rho$ (kg/m³) is air density; $C_D$ is the drag coefficient; $C_h$ is the altitude correction factor; $A_f$ (m²) is the front area; $g$ (m/s²) is gravitational acceleration; $C_r$, $c_1$ and $c_2$ are rolling resistance parameters associated with tire and road conditions; $G$ is road grade.

*C. Fuel Consumption Model*

The microscopic driving model will provide longitudinal vehicle dynamics that can be used to compute fuel consumption based on instantaneous vehicle power demand as in (5) [19],

$$f(t) = \begin{cases} \alpha_0 + \alpha_1 P(t) + \alpha_2 P(t)^2 & P(t) > 0 \\ \alpha_0 & P(t) < 0 \end{cases}, \quad (5)$$

where $f$ is the fuel consumption rate (L/s); $\alpha_0$, $\alpha_1$, $\alpha_2$ are calibrated parameters to be determined; $P$ (kW) is instantaneous vehicle power; $t$ is time. The calculation of model parameters is introduced as follows. Firstly, the vehicle idling fuel consumption is calculated as (6),

$$\alpha_0 = \frac{P_{mfo}\omega_{idle}d}{22164 \times Q_{heat} N_{cyl}}, \quad (6)$$

where $P_{mfo}$ (Pa) is the idling fuel mean pressure; $\omega_{idle}$ (rpm) is the engine idle speed; $d$ (L) is the engine displacement; $Q_{heat}$ (J/kg) is the lower heating value of fuel, and $N_{cyl}$ is the number of engine cylinders. With $\alpha_0$ calculated as in (6), the remaining parameters $\alpha_1$ and $\alpha_2$ are obtained by solving a set of linear equations as in (7),

$$\begin{aligned} F_{city} &= T_{city}\alpha_0 + P_{city}\alpha_1 + P_{city}^2\alpha_2 \\ F_{highway} &= T_{highway}\alpha_0 + P_{highway}\alpha_1 + P_{highway}^2\alpha_2 \end{aligned}, \quad (7)$$

where subscripts *city* and *highway* represent EPA city and highway driving cycles; $F$ (L) is the accumulated fuel consumed in the driving cycle; $T$ (s) is the time duration of the driving cycles; $P$ (kW) is the sum of power in the driving cycles. Such information is publicly available from automotive manufacturers [20]. More details on the model calibration and experimental validation results can be found in [21].

**Remark 1:** It is acknowledged that the vehicle fuel consumption is influenced by additional factors other than instantaneous power demand, such as powertrain configuration, engine operation, auxiliary systems, etc. However, to fairly evaluate the impacts of driving behaviors on vehicle fuel economy, such additional factors should be considered as controlled variables, therefore kept the same during all numerical investigations. As a result, the influences on vehicle fuel economy from controlled variables do not interfere with the principal analysis of driving behaviors. Meanwhile, despite the seemingly trivial formulation, the fuel consumption model is shown to have high fidelity supported by experimental validation from in-field tests [21], which provides a good balance between accuracy and computational demand for the proposed investigation.

### III. SIMULATION CASE STUDY

In this section, simulation cases utilizing the collected commuting traffic data will be designed to reveal the interconnection between the individual driving preferences and vehicle fuel economy, considering various traffic scenarios and vehicle types.

*A. Scenario Setup*

We first collected two sets of daily commuting data with in-vehicle GPS over the period of one week. The geographical information of the routes is shown in Figure 2 and Figure 3. The elevation along each route demonstrates a mostly flat terrain. It is shown that the driver routinely adopts the local route and highway route in the morning and evening, respectively. The commuting data agree with our assumption that the driver tends to follow specific routes during daily operation. Moreover, due to the fixed transportation infrastructures along routes (traffic light, stop sign, etc.) and periodical traffic flow during rush hours, the vehicle speed trajectories demonstrate fluctuations with similar patterns, which can be observed from a sample of collected speed trajectories in the temporal and spatial domains within one-week commute route as in Figure 4 and Figure 5. As expected, the highway and local commute routes have distinctively different speed patterns. Local driving often involves low-speed limits and frequent stops, while highway driving features high cruising speed with fewer speed fluctuations, as shown in Figure 6. In the following context, we use these two trajectories (Day 4) to represent the local and highway commute traffic, respectively.

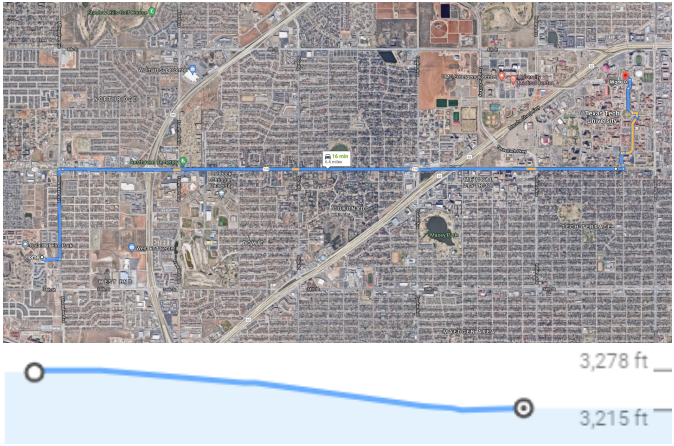

Figure 2. Local route (top, 6.6 miles) with elevation (bottom)

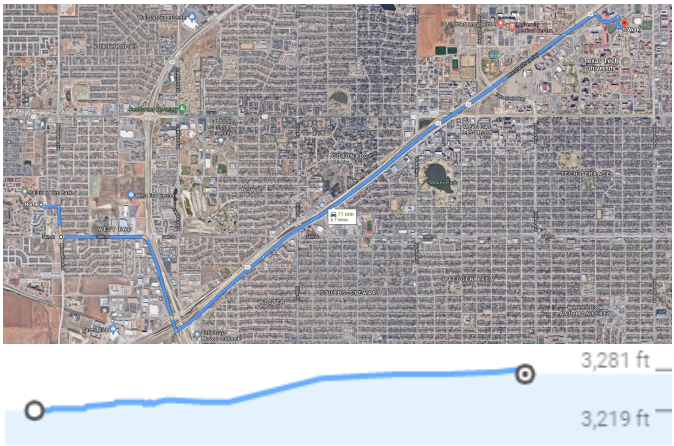

Figure 3. Highway route (top, 6.7 miles) with elevation (bottom)

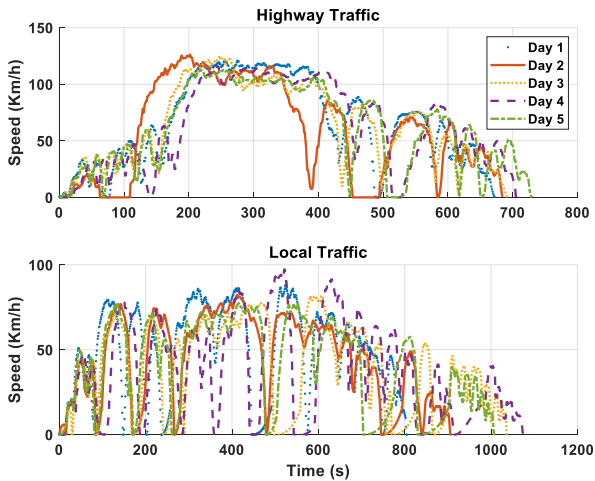

Figure 4. Speed-time trajectories from one week of commute

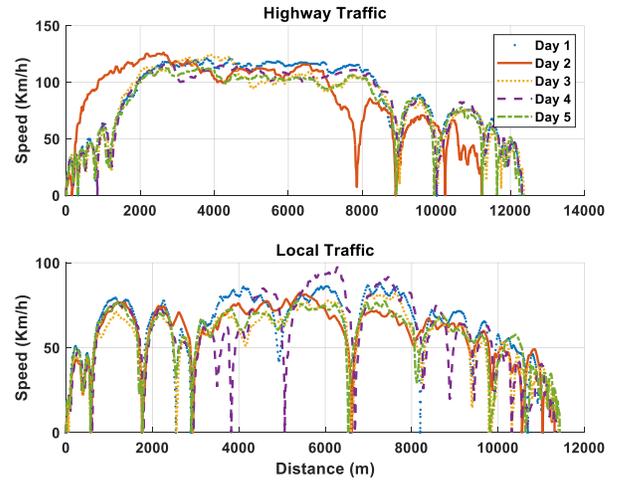

Figure 5. Speed-distance trajectories from one week of commute

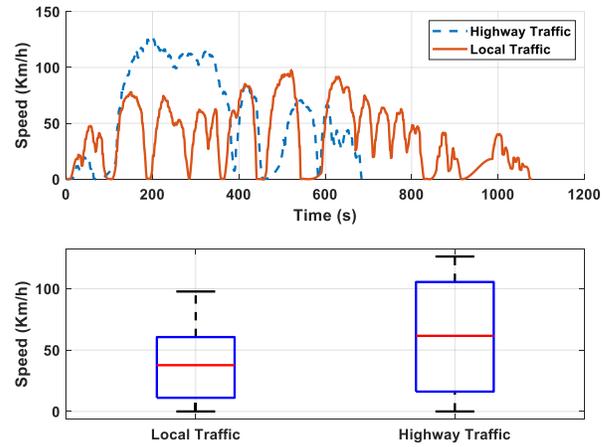

Figure 6. Speed comparison of local and highway traffic

The commute speed trajectories are used as input to the IDM model to represent the speed of the preceding traffic, and the driver will follow the preceding traffic based on the dynamics described in (1) and (2). Depending on drivers' preferred behaviors, the actual speed trajectories vary even under the same traffic conditions. The driving preferences are captured by four parameters encoded in the IDM model, including maximum acceleration, desired deceleration, time headway, and jam distance. These parameters quantitatively characterize how drivers' microscopic behavioral preferences affect vehicle longitudinal dynamics. Among the general population, a range of these parameters has been identified [23] that can replicate realistic and diverse driving behaviors. As shown in Table 1, the differences in driving preference parameters can vary as much as ten times within the realistic range. It implies that driving behaviors among drivers can be dramatically different. With the proposed model, we can efficiently generate diverse driving behaviors in naturalistic traffic conditions. This is particularly beneficial compared with methods that exclusively rely on human subject tests, which are often limited by participants pool, experimental resources, potential safety hazards, and legal issues, leading to insufficiently rich trajectory datasets that may contain biases.

Table 1. Driving preference parameters

| Parameters Category | Realistic Range |
|---|---|
| Maximum Acceleration $a$ (m/s$^2$) | 0.2-2 |
| Desired Deceleration $b$ (m/s$^2$) | 1-3 |
| Time Headway $T$ (s) | 0.8-2 |
| Jam Distance $s_0$ (m) | 1-3 |

In addition to traffic and drivers, vehicle types also affect fuel consumption due to different powertrain and aerodynamics designs. In this study, two representative fuel consumption models are calibrated for the passenger car (Honda Accord) and light-duty truck (Ford F150), as described in section II. The key parameters of the models are shown in Table 2.

Table 2. Fuel consumption model calibration results

| Calibration Parameters | Passenger Car | Light-duty Truck |
|---|---|---|
| Mass (kg) | 1453 | 3152 |
| Drag Coeff. | 0.3 | 0.6 |
| Frontal Area (m$^2$) | 2.32 | 3.87 |
| Cylinders NO. | 4 | 8 |
| Engine Size (L) | 2.4 | 6.2 |
| US City Fuel (mpg) | 22 | 12 |
| US Hwy Fuel (mpg) | 31 | 16 |
| $\alpha_0$ | 5.9217e-4 | 7.7984e-4 |
| $\alpha_1$ | 4.2378e-5 | 1.9556e-5 |
| $\alpha_2$ | 1e-6 | 1e-6 |

*B. Analysis Methodology*

To quantitatively investigate the relationship between driving behaviors and vehicle fuel economy, we first need to identify which driving preference can significantly influence fuel consumption. To this end, a sensitivity analysis of the parameters is essential. Existing sensitivity analysis approaches [24], [25] can be categorically classified as local approaches and global approaches. Local approaches have been proposed first historically, which studies the impact of small perturbations of input around a nominal value on the model output. The foundation of this method relies on calculating or estimating the partial derivative of the model at the nominal point. For accurate and straightforward analytic models, such an approach provides a quick and easy assessment of the sensitivity of parameters at a certain point. However, as the model complexity grows, the method becomes infeasible as the partial derivative is challenging to compute. Moreover, the local approach is only valid at the nominal point where the partial derivative is evaluated, which ignores the possible global variations of the output caused by the input perturbations. It also ignores the potential coupling effects when the model has multiple inputs since the local approach is a one-at-a-time method. To avoid such limitations of local approaches, global sensitivity analysis methods will be adopted in this study. Existing global approaches range from simple screening and scatter plots to the comprehensive decomposition of variance. Considering the aspects of result accuracy, required assumption, and computational effort, the method of importance measures is used to determine the sensitivity of parameters on the model outputs. The method is introduced as follows.

An iterative Monte Carlo simulation is first conducted. At the beginning of each iteration, a set of driving preference parameters $W = (a, b, s_0, T)$, as defined in (1) and (2), is randomly generated within the realistic range, as shown in Table 1. This set of parameters defines the driving behaviors of a driver, which are affected by many factors such as driver demography, traffic conditions, travel purposes [26]. The generated driving preference parameters are used in the microscopic driving model to calculate the vehicle trajectories subject to the preceding traffic. The vehicle trajectories are used to calculate fuel consumption as in (5). At the end of each iteration, the driving preference parameters and corresponding trip fuel consumption will be recorded. This procedure is repeated until the iteration number exceeds a predefined threshold. We require that all trajectories have approximately equal travel time and distance. This requirement is to mitigate the impacts of traffic uncertainties, such as accidents, detours, or overly conservative drivers that lead to prolonged travel time, on the correlation between driving preferences and fuel consumption.

The vehicle speed trajectories from all iterations during local and highway traffic are shown in Figure 7 and Figure 8. The red lines represent the preceding traffic, and the grey shaded areas represent the generated trajectories from 1,000 simulations with diverse driving preferences. As shown in the results, the actual vehicle speed trajectories are profoundly different due to the differences in driving preferences, even in the same traffic scenarios with approximately equal trip distances, as shown in Figure 9. This further underscores the significance of explicitly addressing driving behavioral diversity in naturalistic traffic scenarios.

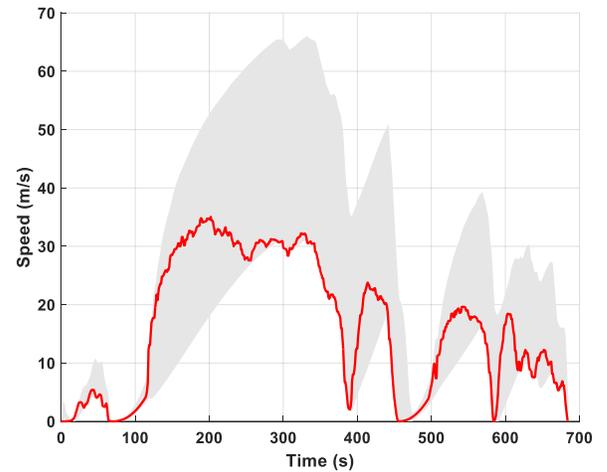

Figure 7. Speed-time trajectories from 1,000 simulations (highway traffic)

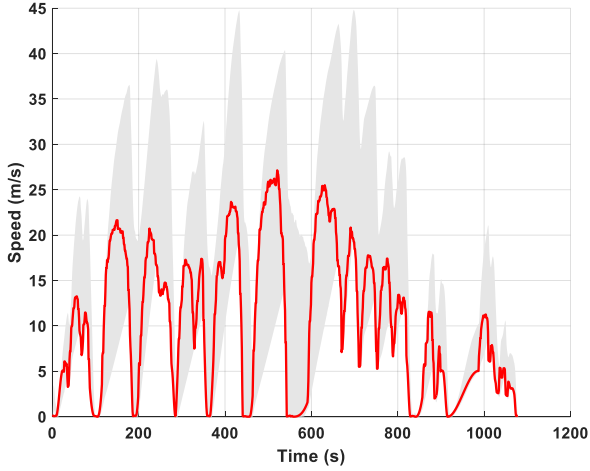

Figure 8. Speed-time trajectories from 1,000 simulations (local traffic)

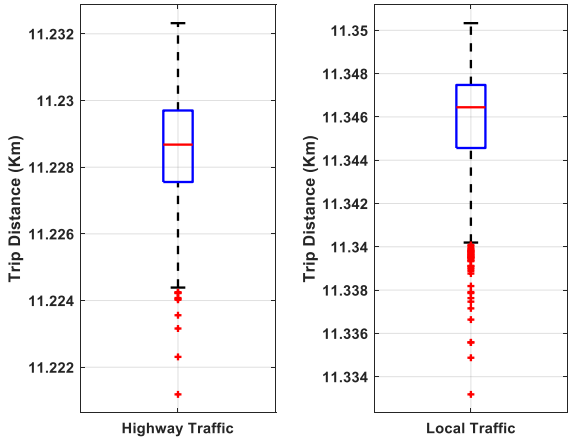

Figure 9. Trip distances from 1,000 simulations

Following the Monte Carlo simulation, we will now measure the importance of each driving preference parameter on fuel consumption by calculating the distance correlation coefficient. The distance correlation coefficient [27] between each driving preference parameter and fuel consumption is calculated as in (8),

$$r_{xy} = \frac{d\operatorname{cov}(X,Y)}{\sqrt{d\operatorname{cov}(X,X)d\operatorname{cov}(Y,Y)}}, \quad (8)$$

$$\begin{aligned}d\operatorname{cov}^2(X,Y) :=& E(\|X-X'\|\|Y-Y'\|) + E(\|X-X'\|)E(\|Y-Y'\|), \\&- E(\|X-X'\|\|Y-Y''\|) - E(\|X-X''\|\|Y-Y'\|)\end{aligned} \quad (9)$$

where $r_{xy}$ is the distance correlation coefficient and $|r_{xy}| \leq 1$; $X$ and $Y$ represent two variables whose correlation are to be determined; $d\operatorname{cov}$ is the distance covariance between two variables; $E$ denotes expected value; A primed variable $X'$ denotes an independent and identically distributed (i.i.d.) copy of an unprimed variable $X$. The distance correlation coefficient is a measure of correlation strength between two vectors where $r_{xy} = 0$ represents no correlation and $|r_{xy}| = 1$ represents a linear correlation between $X$ and $Y$.

---

**Algorithm 1: distance correlation coefficients calculation**
**Input:** preceding traffic speed trajectories
**Output:** distance correlation coefficients

1: Set preceding traffic speed trajectories as $v_{i,i=1}(t)$
2: **FOR** $1 \leq j \leq j_{\max}$
3:     Initialize vehicle speed and acceleration $(v^j_{i,i=2}, \dot{v}^j_{i,i=2}) \leftarrow (0,0)$
4:     Define the $j$th driving reference parameters set as $W_j = (a^j, b^j, s_0^j, T^j)$
5:     Parameterize $W_j \sim U(W_{j,\min}, W_{j,\max})$ where $W_{j,\min}$ and $W_{j,\max}$ are defined according to Table 1
6:     Input $W_j$ to the IDM model (1) and (2) to calculate vehicle speed and acceleration trajectories $(v^j_{i,i=2}(t), \dot{v}^j_{i,i=2}(t))$
7:     Input $(v^j_{i,i=2}(t), \dot{v}^j_{i,i=2}(t))$ to vehicle longitudinal dynamics model (7) and (6) to calculate resistance $R^j(t)$ and power $P^j(t)$
8:     Input power $P^j(t)$ to fuel consumption model (3) to calculate fuel rate $f^j(t)$, then the total trip fuel consumption is $\sum f^j(t)$
9:     Update $j \leftarrow j+1$ and go to step 2
10: **END**
11: Calculate distance correlation coefficient $r_{xy}$ between $\sum f^j(t)$ and individual parameters defined in $W_j$

---

Compared with commonly adopted Pearson correlation, the most distinct advantage of distance correlation is the capability of characterizing the nonlinear relationship of arbitrary dimension variables. The calculation of distance correlation coefficients is summarized in Algorithms 1.

The calculated distance correlation coefficients are shown in Table 3. The value in each cell represents the correlation strength between the corresponding parameter and fuel consumption. Clearly, the larger the coefficients are, the stronger the correlation strength is. Furthermore, we highlight the coefficients that pass a significance test with *p*-values being less than 0.05, indicating high-level confidence in the correlation results. For instance, we find that maximum acceleration and time headway are significantly correlated for both vehicle types in highway traffic. The result indicates that vehicle fuel consumption during highway driving is strongly influenced by driving preferences to apply acceleration and preview the traffic ahead. While in the local traffic, in addition to maximum acceleration and time headway, the desired deceleration can also affect fuel economy. We will discuss these results in detail in the following section.

Table 3. Distance correlation coefficients $r_{xy}$ ($p<0.05$ in blue)

| Traffic | Parameters | Light-Duty Truck | Passenger Car |
|---|---|---|---|
| Highway | Maximum Acceleration $a$ (m/s$^2$) | 0.8419 | 0.8308 |
| | Desired Deceleration $b$ (m/s$^2$) | 0.0016 | 0.0005 |
| | Time Headway $T$ (s) | 0.0210 | 0.0225 |
| | Jam Distance $s_0$ (m) | -0.0008 | -0.0009 |
| Local | Maximum Acceleration $a$ (m/s$^2$) | 0.4339 | 0.4327 |
| | Desired Deceleration $b$ (m/s$^2$) | 0.0070 | 0.0118 |
| | Time Headway $T$ | 0.0127 | 0.0128 |
| | Jam Distance $s_0$ (m) | 0.0009 | -0.0010 |

**Remark 2**: Selecting the maximum iteration number of Monte Carlo simulations is an empirical task since there is no general and analytical rule to determine the required number of iterations given a fixed confidence interval. Recalling that the purpose of the Monte Carlo simulation is to find the correlation strength between driving preference parameters and fuel consumption, it is proposed that the iteration number deems to be sufficient when the calculated correlation coefficients converge to constant levels as the iteration number increases. The evolution of the correlation coefficient, along with the iteration number, is shown in Figure 10 and Figure 11. There are four lines in each plot, representing four parameters' correlations with respect to fuel consumption in local and highway driving scenarios. After 1,000 runs of simulation, the resulting correlation coefficients converge to constant levels as expected. Therefore, it is determined that the Monte Carlo simulation results are representative.

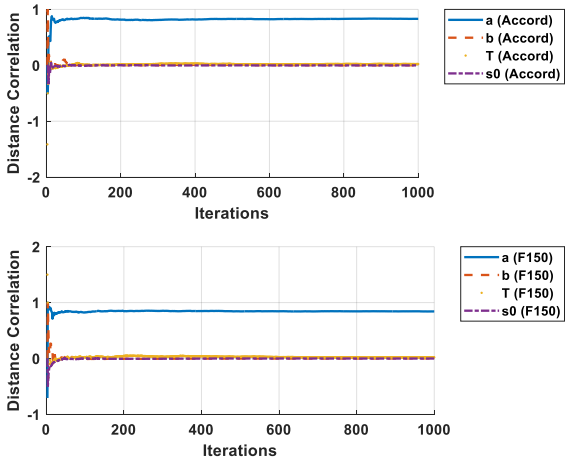

Figure 10. Distance correlation coefficients evolution (highway)

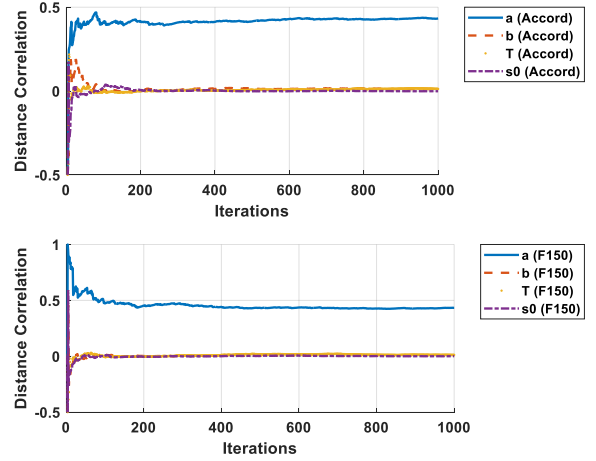

Figure 11. Distance correlation coefficients evolution (local)

*C. Results Discussion*

It can be easily identified from Table 3 that the correlation coefficients of maximum acceleration are at least one order of magnitude higher than any other correlation coefficients in all traffic scenarios and vehicle types. In other words, maximum acceleration has the strongest correlation with fuel economy, therefore, has the potential to serve as a primary predictor. We visualize the correlations between maximum acceleration and fuel consumption in Figure 12, where clear nonlinear correlating patterns can be observed in each scenario.

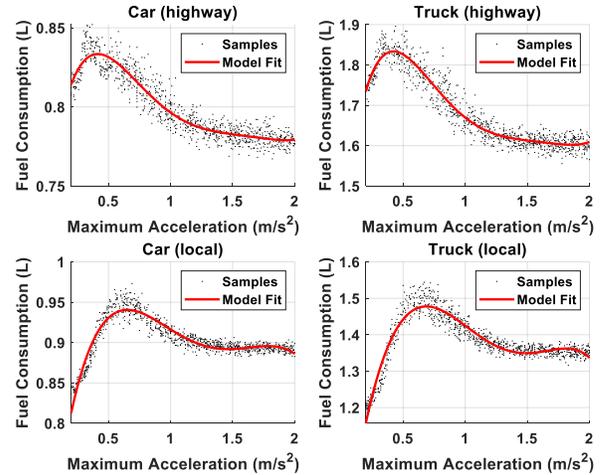

Figure 12. Correlations between maximum acceleration and fuel consumption

Furthermore, such a correlation can be captured by a quintic function as in (10),

$$\zeta(a) = z_1 a^5 + z_2 a^4 + z_3 a^3 + z_4 a^2 + z_5 a + z_6, \quad (10)$$

where $a$ is the maximum acceleration; $z_{1\sim6}$ are fitting constants determined by least-squares regression. The numerical model-fitting results are summarized in Table 4. A low Root Mean Square Error (RMSE) indicates the quintic function is able to capture the correlation with high accuracy, and the high R-squared values indicate the strong capability of the quintic

8model to explain the data variances. Therefore, the fitted quintic function provides an adequate estimation of vehicle fuel consumption based on drivers' preferred maximum acceleration.

Table 4. Maximum acceleration model performance

| Traffic | Vehicle | R-Square | RMSE (L) |
|---|---|---|---|
| Highway | Light-Duty Truck | 0.927 | 0.0235 |
| | Passenger Car | 0.8939 | 0.0067 |
| Transient | Light-Duty Truck | 0.8933 | 0.02267 |
| | Passenger Car | 0.8814 | 0.008529 |

Further inspection of the correlations in Figure 12 reveals more profound insights into the impacts of drivers' maximum acceleration on vehicle fuel consumption. In the local traffic scenario, the lowest fuel consumption occurs when the drivers have the lowest maximum acceleration, around 0.2 m/s$^2$. As the maximum acceleration increases to 0.6 m/s$^2$, the fuel consumption increases almost linearly and peaks around 0.6 to 0.8 m/s$^2$. Then, the fuel consumption decreases as the maximum acceleration increases until 1.2 m/s$^2$, after which the fuel consumption is not significantly affected by the change of maximum acceleration. Both vehicles demonstrate similar fuel consumption trends with respect to the maximum acceleration, whereas the light-duty truck shows a more significant variance of 29% from bottom to peak, compared with 14% for the passenger vehicle. This correlation qualitatively suggests that one may prefer to drive with a lower maximum acceleration to achieve a better fuel economy in local traffic. This finding is practically beneficial for applications such as parcel delivery, which typically adopt light-duty trucks and vans that operate with similar daily local routines.

In the highway traffic, however, we observed distinctively different correlation relationships between the drivers' maximum acceleration and vehicle fuel consumption. When the drivers have low maximum accelerations ranging from 0.2 m/s$^2$ to 0.4 m/s$^2$, a positive correlation is observed until the fuel consumption peaks at 0.4 m/s$^2$. After that, the fuel consumption sharply decreases with the increase of maximum acceleration. As a result, it is found that more fuel-economical driving behaviors in the highway traffic require higher maximum accelerations ranging from 1.4 m/s$^2$ to 2 m/s$^2$. This conclusion is the opposite of local traffic due to the dramatically different traffic patterns, where highway traffic features much fewer stop-and-go scenarios and higher cruise speed, as shown in Figure 6. Similar to the local traffic, the light-duty truck has a larger fuel consumption variance of 23%, while the passenger car has a variance of 15%. Besides, the peak fuel consumption in all four scenarios appears around 0.4 m/s$^2$ to 0.6 m/s$^2$ suggesting a common region of undesirable driving behaviors in terms of fuel economy. To summarize the findings from Figure 12, we confirm the strong and consistent correlations between drivers' maximum acceleration and vehicle fuel consumption with different traffic scenarios and vehicle types. These correlations provide general guidance to estimate vehicle fuel consumption based on the drivers' maximum acceleration.

In addition to the maximum acceleration, other driving preferences, such as time headway and desired deceleration, also have non-negligible influences on vehicle fuel consumption despite their much lower correlation strengths, as shown in Table 3. This suggests that such preferences may not be used to single-handedly predict fuel consumption but can potentially serve as supplemental indicators to enhance the accuracy of the prediction model (10). Driven by this motivation, we will present a more rigorous model to characterize the correlation quantitatively and evaluate its performance enhancement numerically in the following context.

The new prediction model takes the driving preference parameters as inputs and predicts vehicle fuel consumption as output. We utilize a Gaussian Process Regression (GPR) model to train and predict the fuel consumption of drivers operating on frequent routes. A GPR model is a nonparametric kernel-based probabilistic model that predicts the value of a response variable $y_{new}$, given the input features vector $x_{new}$ and the training data set $(X,Y) = \{(x_i, y_i) | i = 1, 2, \cdots n\}$. The input features are selected according to Table 3 where $p < 0.05$. The GPR model takes the form of (11),

$$y = h(x)^T \beta + g(x), \quad (11)$$

$$g(x) \sim GP(0, k(x, x')), \quad (12)$$

$g(x)$ are latent variables from a Gaussian process with zero mean $E(g(x)) = 0$ and covariance function,

$$k(x, x') = Cov[g(x), g(x')] = E[g(x)g(x')]. \quad (13)$$

$h(x)$ are a set of basis functions that transform the original feature vector into a new feature vector in $R^m$. $\beta$ is an $m$-by-1 vector of basis function coefficients. The covariance function $k(x, x')$ of the latent variables $g(x)$ captures the smoothness of the response, and basis functions $h(x)$ project the inputs into an $m$-dimensional feature space. Response $Y$ can then be written as,

$$P(Y|G, X) \sim N(Y|H\beta + G, \sigma^2 I), \quad (14)$$

$$\begin{aligned}X &= (x_1^T, x_2^T, \cdots x_n^T)^T \\ Y &= (y_1, y_2, \cdots y_n)^T \\ H &= (h(x_1^T), h(x_2^T), \cdots h(x_n^T))^T, \\ G &= (g(x_1), g(x_2) \cdots g(x_n))^T\end{aligned} \quad (15)$$

and $\sigma^2$ is the noise variance. The covariance function $k(x, x')$ is parameterized by a set of kernel functions. In this study, we adopt the Matern 5/2 kernel defined as

$$k(x, x') = \sigma_f^2 \left(1 + \frac{\sqrt{5}r}{\sigma_l} + \frac{5r^2}{3\sigma_l^2}\right) \exp\left(-\frac{\sqrt{5}r}{\sigma_l}\right),$$

$$r = \sqrt{(x-x')^T (x-x')} \quad (16)$$

where $\sigma_f$ is the standard deviation and $\sigma_l$ is the characteristic length scale that defines how far apart the input $x$ can be for the response to become uncorrelated. Both $\sigma_f$ and $\sigma_l$ are positive and can be parameterized by a vector $\theta$ such that,

$$\begin{aligned}\theta_1 &= \log \sigma_f \\ \theta_2 &= \log \sigma_l\end{aligned}. \quad (17)$$

The estimation of parameters $\hat{\beta}, \hat{\theta}, \hat{\sigma}$ is conducted by maximizing the likelihood $P(Y|X)$ as a function of $\beta, \theta, \sigma$ over the training data set,

$$\hat{\beta}, \hat{\theta}, \hat{\sigma} = \arg\max_{\beta,\theta,\sigma} \log P(Y|X,\beta,\theta,\sigma), \quad (18)$$

where the marginal log-likelihood function is,

$$\begin{aligned}\log P(Y|X,\beta,\theta,\sigma) = \\ -\frac{1}{2}(Y-H\beta)^T \left(K(X,X)+\sigma^2 I_n\right)^{-1}(Y-H\beta). \\ -\frac{n}{2}\log 2\pi - \frac{1}{2}\log\left|K(X,X)+\sigma^2 I_n\right|\end{aligned} \quad (19)$$

Here the covariance functions for the joint distribution of latent variables are denoted as

$$K(X,X) = \begin{pmatrix} k(x_1,x_1) & k(x_1,x_2) & \cdots & k(x_1,x_n) \\ k(x_2,x_1) & k(x_2,x_2) & \cdots & k(x_2,x_n) \\ \vdots & \vdots & \vdots & \vdots \\ k(x_n,x_1) & k(x_n,x_2) & \cdots & k(x_n,x_n) \end{pmatrix}. \quad (20)$$

The solution of (18) yields the GPR model for prediction, which estimate the expected value of prediction $y_{new}$ at $x_{new}$ over the validation data set,

$$\begin{aligned}E(y_{new}|Y,X,x_{new},\beta,\theta,\sigma) = \\ h(x_{new})^T \beta + \\ K(x_{new}^T,X)\left(K(X,X)+\sigma^2 I_n\right)^{-1}(Y-H\beta)\end{aligned} \quad (21)$$

The detailed proof can be found in [28]. The prediction models are cross-validated with five folds under each of the driving scenarios. The validation results are shown in Table 5. Compared with the quintic model in Table 4, the GPR model surpasses the performance measurements in all scenarios considered.

Table 5. GPR model performance

| Traffic | Vehicle | R-Square | RMSE (L) |
|---|---|---|---|
| Highway | Light-Duty Truck | 0.95 | 0.018807 |
| | Passenger Car | 0.92 | 0.0058608 |
| Transient | Light-Duty Truck | 0.94 | 0.017209 |
| | Passenger Car | 0.94 | 0.0062848 |

### D. Personalized Driving Application

A practical application of the prediction model is to develop a personalized fuel-friendly driving strategy for drivers who operate frequent routes on a daily basis, such as commuting, public transportation, and delivery. As drivers regularly drive over frequent routes, the vehicle can collect the driving data, including speed, acceleration, and fuel consumption measurement. The speed and acceleration trajectories can be used to estimate the drivers' preferences, such as the maximum acceleration, desired deceleration, and time headway, assuming the vehicle is equipped with proper sensors. This assumption does not necessarily introduce additional hardware modifications or costs since many mass-production vehicles are already equipped with onboard radar and cameras nowadays. The driving preferences and fuel consumption can then be used as training and validation datasets for the prediction model to learn drivers' behaviors and predict the vehicle fuel economy. As drivers accumulate more miles on the frequent routes due to daily driving, the training and validation datasets expand accordingly, which improves the model's adaptability of tailoring to specific human drivers' preferences. Based on the model prediction, the vehicle can then suggest drivers adopting specific routes and driving behaviors that will lead to reduced fuel consumption without incurring prolonged trip duration. The structure of the proposed implementation is shown in Figure 13.

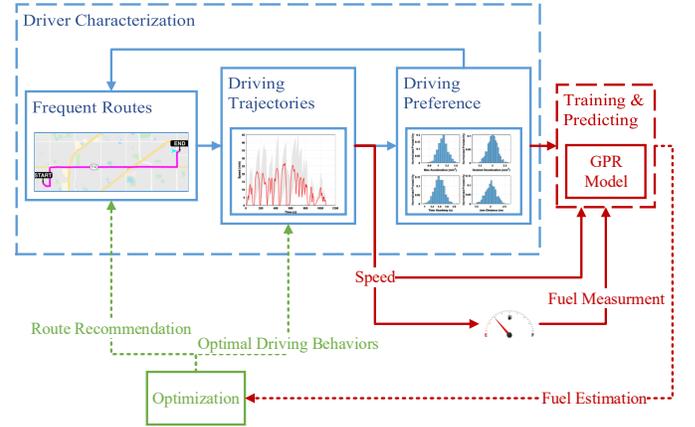

Figure 13. Personalized fuel friendly driving strategy

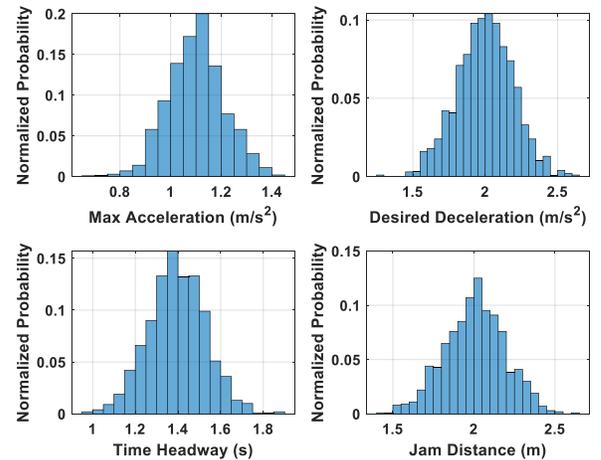

Figure 14. Test driving preference parameters distributions

To demonstrate the proposed design concepts and verify the generalizability of the trained model, we further conduct a model test using a new dataset of driving preference parameters, including 1,000 samples that are not used in previous training and validation datasets, as shown in Figure

14. The *x*-axis represents the range of each driving preference parameter, and the *y*-axis represents the normalized probability. The test driving preference parameters are generated from a normal distribution with a mean value equal to the average of the realistic range as in Table 1, and the standard deviation is 10% of the mean value, as shown in (22),

$$W_{test} \sim N\left(\bar{W}_{test}, \left(0.1 \times \bar{W}_{test}\right)^2\right). \quad (22)$$

The practical implications of normal distributions in driving preference parameters are based on the assumption that an individual's driving behaviors are influenced by not only his or her intrinsic characteristics but also the uncertain external environment (travel purpose, weather condition, etc.). Therefore we consider the individual has driving behaviors governed by a normal distribution where the mean value represents the intrinsic driving preference, and the standard deviation represents the environmental uncertainty. In practice, different drivers have various distributions of driving preferences, which can be estimated [29] via continuously collected driving data on frequent routes.

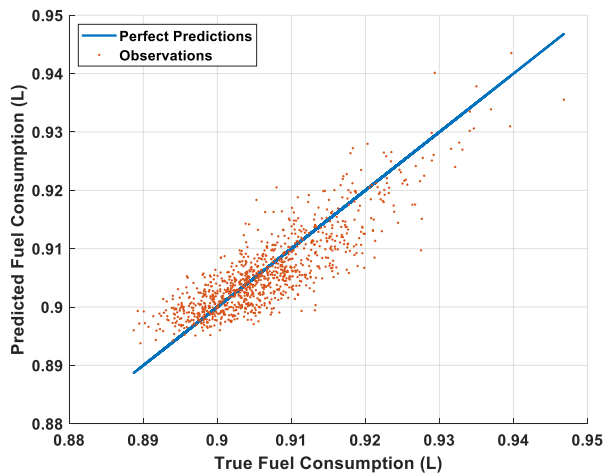

Figure 15. Fuel consumption prediction (model test)

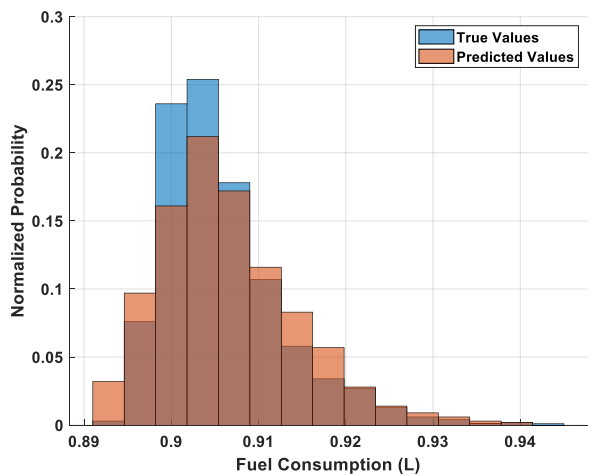

Figure 16. Predicted fuel consumption distribution (model test)

We visualize the test result for a passenger car operating under transient traffic as an example. As shown in Table 3, in the local traffic, the maximum acceleration, desired deceleration, and time headway are the driving preferences that have substantial correlations with vehicle fuel consumption. Hence these three features are used as the inputs for the trained GPR model to predict fuel consumption. The exact prediction results can be found in Figure 15. The solid blue line represents the perfect prediction, and the red points cloud represents observed fuel consumption. The prediction error is 0.0043 L (RMSE). The test result shows the model's accuracy and generalizability to predict fuel consumption based on the correlations developed earlier.

A statistical comparison result of the GPR model against the ground truth is shown in Figure 16, where the *y*-axis is the normalized probability. Practically, this can be interpreted as the estimated likelihood of fuel consumption for the individual driving on specific routes. To evaluate the model accuracy from this statistical perspective, we need to quantitatively measure the similarity between the predicted distribution $Q$ and the true distribution $P$ by Kullback–Leibler (KL) divergence, as calculated in (23),

$$D_{KL}(P \| Q) = \sum_{i=1}^{N} P(x_i) \log\left(\frac{P(x_i)}{Q(x_i)}\right), \quad (23)$$

where $x_i$ represents the event for each discretized probability distribution, and $N$ is the total number of events. It can be shown that $D_{KL}(P \| Q) \in [0, \infty)$ where the lower value of KL divergence represents a high similarity between two distributions and vice versa [30]. The test result has a KL divergence of 0.0823, indicating a good match between the predicted distribution and ground truth. The comparison results in Figure 15 and Figure 16 collectively demonstrate the quantitative impacts of drivers' behaviors on the fuel consumption of vehicles operating on the frequent routes in a statistically significant manner.

## IV. CONCLUSIONS AND FUTURE WORK

The impacts of various driving behaviors on vehicle fuel economy in different traffic scenarios are quantitatively investigated in this study. This is conducted through a combined Monte Carlo simulation and experimentally collected naturalistic traffic trajectories, which allows us to efficiently evaluate diverse and realistic driving data that are otherwise difficult to obtain through in-field tests. Results suggest that drivers' preferred maximum acceleration plays the most critical role when determining vehicle fuel consumption on frequent routes. Such a relationship can be described by a quintic polynomial that is capable of capturing the primary nonlinear correlation between fuel consumption and maximum acceleration. Additional findings reveal that driving behaviors like desired deceleration and time headway also have statistically significant impacts on fuel consumption in local and highway traffic, despite much weaker correlations compared with maximum acceleration. By identifying these significant behavioral factors that affect fuel consumption, we can utilize these parameterized driving preferences as featured inputs to train a Gaussian Process Regression model that can predict the vehicle fuel consumption with high accuracy, generalizability, and efficiency. The model can be used to estimate vehicle fuel economy based on the drivers' preferences over frequent routes, recommend fuel-friendly routes and driving strategies, and better accommodate individual driving needs from daily operation.

Future studies, including a personalized eco-driving strategy for the automated vehicle to improve traffic energy efficiency, and real-time driving behaviors modeling, will be expanded upon the insights gained in this study.

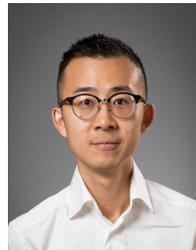

**Yao Ma (M'19)** received the BE degree in Control Science and Engineering from Harbin Institute of Technology, Harbin, China, in 2012; the MS degree in Electrical and Computer Engineering from North Carolina State University, Raleigh, NC, in 2014; and the Ph.D. degree in Mechanical Engineering from the University of Texas at Austin, Austin, TX, in 2019.

Dr. Ma is currently an Assistant Professor in the Department of Mechanical Engineering at Texas Tech University. His research interests focus on the control and modeling of intelligent vehicle systems for the improvement of efficiency, mobility, and safety.

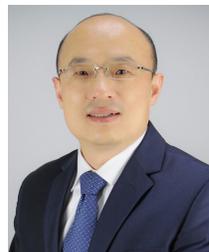

**Junmin Wang (SM'14)** received the BE degree in automotive engineering and the MS degree in power machinery and engineering from Tsinghua University, Beijing, China, in 1997 and 2000, respectively; the second and third MS degrees in electrical engineering and mechanical engineering from the University of Minnesota, Twin Cities, MN, USA, in 2003; and the Ph.D. degree in mechanical engineering from the University of Texas at Austin, Austin, TX, USA, in 2007.

Dr. Wang is the Lee Norris & Linda Steen Norris Endowed Professor in the Walker Department of Mechanical Engineering at the University of Texas at Austin. His major research interests include control, modeling, estimation, and diagnosis of dynamical and mechatronic systems. Prof. Wang is an IEEE Vehicular Technology Society Distinguished Lecturer (2015-2019), SAE Fellow (2015), and ASME Fellow (2016).